\documentclass[]{elsart}
\journal{Thin Solid Films}
\usepackage{amssymb}
\usepackage{graphicx}
\usepackage[english]{babel} 

\begin{document}
\begin{frontmatter}

\title{Quenched growth of nanostructured lead thin films on insulating substrates}

\author[MSU]{V.E. Bochenkov\corauthref{cor}},
\ead{boch@kinet.chem.msu.ru}
\corauth[cor]{Corresponding author. Tel.: +7 (095) 939-5442, Fax.: +7 (095) 939-0283.}
\author[PKM]{P. Karageorgiev},
\author[PKM]{L. Brehmer} and 
\author[MSU]{G.B. Sergeev} 

\address[MSU]{Department of Chemistry, Lomonosov Moscow State University, 119992 Moscow, Russia}
\address[PKM]{Institut f\"ur Physik, Universit\"at Potsdam, Am Neuen Palais 10, 14469 Potsdam, Germany}

\begin{abstract}
Lead island films were obtained via vacuum vapor deposition on glass and ceramic substrates at 80~K. Electrical conductance was measured during vapor condensation and further annealing of the film up to room temperature. The resistance behavior during film formation and atomic force microscopy of annealed films were used as information sources about their structure. A model for the quenched growth, based on ballistic aggregation theory, was proposed. The nanostructure, responsible for chemiresistive properties of thin lead films and the mechanism of sensor response are discussed.
\end{abstract}

\end{frontmatter}

% main text
\section{Introduction}
\label{introduction}
Thin films attract significant attention of the scientists not only due to the growing technological applications, but also because of the interest in confinement effects which arise with decreasing of one or more dimensions of the material~\cite{Meiwes}. The properties of such systems are dramatically influenced by their nanoscale structure. The growth of vacuum deposited films can occur by one of three mechanisms: two-dimensional (2D) layer-by-layer growth (Frank-van-der-Merwe type), three-dimensional (3D) growth (Volmer--Weber type), or mixed type, with 2D film formation followed by 3D island growth (Stranski--Krastanov type)~\cite{Zinke92}. The thermodynamically preferable mode is defined by the surface free energies of the substrate and film and also their interfacial free energy. If the difference between surface free energies of the clean substrate in vacuum and that of the clean material exceeds the interfacial free energy, the material doesn't wet the substrate and 3D structures are grown. Usually this is the case for metals on oxide substrates~\cite{Campbell97}. The formation of 3D structure is often considered undesirable, since it leads to rough, porous films~\cite{SmithBook}. However, this kind of structure can play a key role in the gas sensor technology. 

Recently, the gas sensitivity to ammonia and humidity has been found for cryochemically synthesized lead-containing poly(\emph{p}-xylylene) films~\cite{patent,analcomm} and lead thin films deposited on poly(\emph{p}-xylylene) layer~\cite{Bochenkov02}. In order to distinguish the influence of different factors on their sensitivity, the exclusion of the polymer from the system under investigation is of particular interest. The present work has been inspired by recently found chemiresistive properties of thin lead films grown directly on ceramic substrates at 80~K. The focus of this investigation is to study the growth of lead films on cold insulating substrates and their resulting structure, which is responsible for the observed sensitivity.

\section{Experimental details}
\label{experimental}
Lead of 99.99~\% purity was evaporated from Ta boat at the temperature in range 540--680
$^\circ$C by resistive heating. Substrates were mounted on stainless steel substrate holder, which allowed cooling down to 80~K by liquid nitrogen flow, so that the incident angle was 0$^\circ$ (perpendicular to the substrate). Two J-type thermocouples with Voltcraft 302~K/J digital thermometer were used to control evaporation and substrate temperatures. The experiments were performed at residual pressure $\leq{10^{-6}}$ Torr, with the use of rotary and turbomolecular pumps. The film deposition rate and thickness were monitored by quartz balance FTM~7 (BOC Edwards, UK). 

The resistance of the films was measured \emph{in situ} using Keithley 617 Electrometer. The voltage in range 1--10 Volts from the built-in source was applied to the substrate and the current was measured. No distinction was found between the curve shapes obtained at different voltages in this range. The effective resistance $R_{eff}$ was calculated by dividing the voltage applied on the current measured without verification of the Ohm's law fulfillment.

Atomic force microscopy (AFM) investigations have been carried out with Autoprobe CP microscope (Park Scientific Instruments, USA) operating in low amplitude oscillating mode at room temperature in the air. Commercial Si cantilevered tip was used as a probe. The curvature radius of the probe tip (31 nm) was calibrated using Au-nanoparticles standard kit in order to get proper information about the dimension of the nanosized Pb islands and interisland distances. 

Typical experiment involved evacuation of the chamber, cooling the substrate down to 80--130~K, deposition of lead, and annealing of the film up to the room temperature with the rate $<0.3$~K/sec. The resistance of the sample was recorded during the deposition and heating. Two types of substrates were used: ceramic substrates with Ni-Cr interdigitated electrodes and glass substrates with straight pre-deposited Au/Cr electrodes. No difference was observed for resistance behavior of the films prepared on these substrates. The samples on the glass were investigated by AFM after exposition to the air. 

Electrical measurements at different humidity were carried out using Keithley 617 electrometer in a special testing chamber, which allowed fast changes of the atmosphere by switching between two gas flows. The dry and humidified nitrogen was used for measuring the change of the current through the film with the change of relative humidity in the chamber.

\section{Results}
\label{results}
\subsection{Resistance of the film during the deposition}
\label{deposition}
The resistance of the pure substrates was in range $2\cdot10^{10}$--$5\cdot10^{11}$ Ohm. At the beginning of the deposition resistance remains constant and doesn't change until the critical concentration of Pb on the surface is reached. Continuation of the condensation above this amount causes the decrease of resistance. Finally, it leads to the formation of an infinite conducting cluster. In present work this critical concentration was found to be dependent on the deposition rate, which is varied via evaporation temperature. The lower the temperature, the higher the critical concentration is. For instance, for the sample, deposited at $6.6\cdot10^{13}$ atoms/sec$\cdot$cm$^2$ and contained $2.8\cdot10^{16}$ atoms/cm$^2$, no conductance onset has been observed, but for the rate of $1.4\cdot10^{14}$ atoms/sec$\cdot$cm$^2$ it has occurred at $2.0\cdot10^{16}$ atoms/cm$^2$, and for the films condensed at the rate of $3.0\cdot10^{14}$ atoms/sec$\cdot$cm$^2$ the critical concentration has been found to be $1.3\cdot10^{16}$ atoms/cm$^2$. 

The samples were prepared in such a way that the deposition was stopped after the beginning of the resistance decrease, at 10--20~\% change of R$_{eff}$. Thus, the films achieved the percolation threshold were slightly more conductive than the others, which didn't. After the deposition the resistance was constant at constant temperature. 

Thus, it was found that by changing the evaporation temperature one can prepare samples with the same amount of lead with and without percolation at 80~K. 

\subsection{Resistance of the film during the annealing}
\label{annealing}
\begin{figure}[h]
\includegraphics[width=82.5mm]{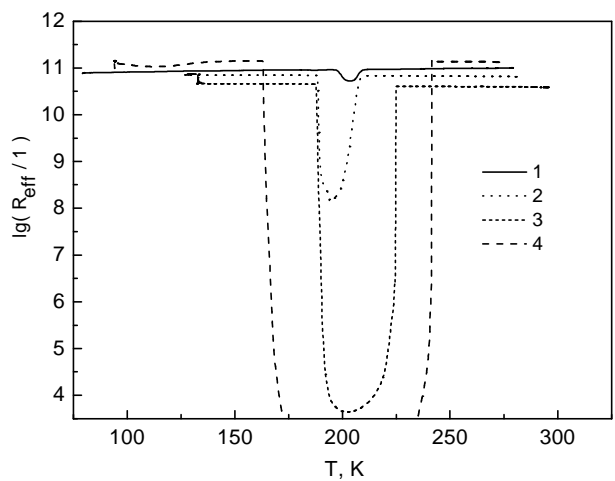}
\caption{Resistance change of lead thin films during the annealing. Experimental details are presented in Table~\ref{tab1}. } 
\label{fig1}
\end{figure}

The samples obtained with various amount of lead were heated to 300~K. During annealing, the minima on the resistance curves were observed for all samples at $200\pm$5~K. In Fig.~\ref{fig1} the typical curves of effective resistance as a function of temperature are presented for the films deposited at different rates. The experimental details for corresponding samples are shown in Table~\ref{tab1}. The different initial resistance values in Fig.~\ref{fig1} are because of the different resistances of the pure samples. The change of the deposition temperature in range 80--130~K (which caused the different initial temperatures in Fig.~\ref{fig1}) revealed no effect on electrical behavior of the films during annealing. One can see that the resistance change during the annealing is larger for the samples prepared at higher deposition rates. 

\begin{table}[h]
\caption{Experimental details for the thin lead film samples, corresponding to the resistance curves, presented in Fig.1.}
\begin{tabular}{|c|c|c|c|}
\hline
Curve & Pb amount/ & Deposition rate/ & Conductance \\
No. in fig.~1  & $10^{15}$ atoms/cm$^2$ & $10^{13}$ atoms/cm$^2\cdot$s  & onset \\
\hline
1 & 6.6 & 3.3 & not observed \\
2 & 28.1 & 4.3 & not observed \\
3 & 22.4 & 6.6 & observed \\
4 & 23.1 & 16.5 & observed \\
\hline
\end{tabular}
\label{tab1}
\end{table}

A series of experiments with sequential heating and cooling cycles have been carried out in order to get more information about the processes occurring in films during annealing. The samples were heated up to a certain temperature, then cooled down to 80~K. Then the heating--cooling cycle was repeated with higher upper temperature limit. 

It was found that at the beginning of annealing the conductance change is fully reversible up to approximately 100~K. Heating increased the current, and cooling returned its value, achieved after the deposition. Obviously, no change of the structure occurred below this temperature. 

After annealing above 100~K the conductance change became irreversible. Cooling to 80~K decreased the current insignificantly. The most conductive structure could be fixed by freezing the sample, previously annealed to 200~K. The films annealed above 240~K, didn't change conductance with the temperature. 

\subsection{AFM and electrical investigations of annealed films}
\label{AFM}
\begin{figure}[h]
\includegraphics[width=82.5mm]{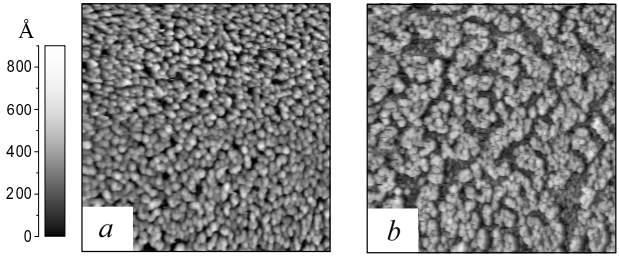}
\caption{AFM images of quench condensed Pb films after annealing to room temperature and exposition to the air. Scan area is $5\mu m\times5\mu m$. (a) No conductance onset during the deposition was observed. (b) The conductance onset was observed.} 
\label{fig2}
\end{figure}

The annealed lead island films were characterized by AFM after exposure to the air. The microstructure of the films was found to be different for the deposits, which didn't achieved the conductance onset, and the others, which did. The former films contain lead nanoparticles uniformly distributed over the surface. The aggregation of the nanoclusters was observed in the latter case. The topographies of two films, containing similar amounts of lead, but prepared at different deposition rates, are depicted in Fig.~\ref{fig2}. Part~\emph{a} corresponds to curve 2 in Fig.~\ref{fig1}, and \emph{b} --- to curve 4 (see the corresponding data in Table~\ref{tab1}). The interisland distances in Fig.~\ref{fig2}\emph{a} are visualized shorter than in the real structure due to distortion caused by finite curvature radius of the tip.

Electrical measurements at different relative humidity in range 0--95~\% at room temperature was shown the reversible sensor response of samples, for which the conductance onset had been achieved during the deposition. The response time was about 1--2 seconds, and the resistance change was of 4--5 orders of magnitude. However, the more detailed investigation of chemiresistive properties of such films is beyond the scope of present article and left for the future experimental studies. 

\section{Discussion}
\label{discussion}
Thin films, grown at $T_s/T_m<0.3$, where $T_s$ and $T_m$ are substrate and melting temperatures, respectively, are usually referred to as quench condensed (QC), and the  region, where this expression is satisfied --- Zone 1 (Z1)~\cite{SmithBook}. The growth of such films is being extensively studied, but the structure-formation mechanism is still under discussion.

The study of resistance behavior of thin lead films deposited at $T_s$=4.2~K during annealing to room temperature was carried out in~\cite{Danilov98}. The minimum found on the curve $R(T)$ is in an agreement with the results obtained in present work. However, no influence of the evaporation temperature on the percolation threshold and the structure of the island films has been reported. The authors note the unusually high amount of lead, required for the conductance onset. The explanation proposed is based on the idea that lead atoms during the deposition at low temperature can form diatomic molecules Pb$_2$, which possess the repulsion potential to the arriving Pb atom. Thus, the metastable continuous amorphous phase is created. However, it was shown both experimentally~\cite{Gingerich76} and theoretically~\cite{Bala01,Zhao02} that lead clusters Pb$_n$ with $n>2$ are energetically stable. 

Another series of investigations have been carried out in order to understand the nature of QC lead films ~\cite{Ekinci98,Ekinci99}. The authors have used \emph{in situ} scanning tunneling microscopy (STM) to study lead deposits on Highly Oriented Pyrolityc Graphite (HOPG) at 77 or 4~K. The results indicate that the films with the thickness $>0.9$~nm have discontinuous structure with regions, covered by lead and separated by voids. The authors consider observed structures as a lead nanocrystallites, formed at low temperature from amorphous phase via avalanche-like crystallization process. 

Taking into account the results of low temperature STM~\cite{Ekinci98,Ekinci99} and resistance measurements~\cite{Danilov98}, and also accounting the found effect of metal evaporation temperature, we proposed a model, which is based on the ballistic aggregation theory~theory assumes the absence of surface diffusion during \cite{SmithBook}. This the deposition, which is realized at low substrate temperature. Besides, the attraction between arriving atom and atoms of growing cluster are taken into account. Thus, if the ratio of incident energy $E_t$ to the adatom potential--well depth $E_p$ is not too high, the particles stick to the cluster, not reaching the substrate. It results in the formation of highly porous film with columnar structure. Such a structure can be predicted by molecular dynamic simulations~\cite{Mueller87}. Obviously, such discontinuous films must be insulating. When the kinetic energy of arriving atoms becomes high enough, they begin to bypass the columns and stick to the substrate, thus filling the voids. The calculation predicts columnar structure growth for the ratio $E_t/E_p$=0.02. At the ratio $E_t/E_p$=0.5 the interconnected columns grow, that would cause the increase of  the film conductance~\cite{Mueller87}. 

The potential-well depth, taken from experimental value of dissociation energy for clusters Pb$_2$, is equal to 81$\pm$6~kJ/mole~\cite{Gingerich76}. Roughly estimating the mean kinetic energy of adatoms evaporated at temperature $T$  in the range of 540--680~$^\circ$C by formulae $E_t=\frac{3}{2}kT$, one can obtain the energy $E_t$=10--12~kJ/mole and the ratio $E_t/E_p\approx$0.12--0.14, that is consistent with the ballistic aggregation model.
 
More thoroughly the influence of different factors, such as substrate temperature, incident angle, and kinetic energy of adatoms on voids formation has been studied, as an example, on nickel films using molecular dynamics~\cite{Zhou97}. In this work the concentration of vacancies was found to decrease exponentially with the increase of adatom energy. This fact is consistent with our preliminary results on the dependence of critical amount of lead, necessary to achieve the conductance onset, \emph{versus} the evaporation temperature.

Within the framework of the ballistic aggregation model the structures observed in STM images obtained at 77 and 4~K~\cite{Ekinci98,Ekinci99} can be considered as a top-view of structures, discussed above. The effect of evaporation temperature on lead concentration, required for conductance onset is supposedly arisen from increasing the ratio $E_t/E_p$ and thus decreasing the voids. 

Further behavior of the film conductance with heating can therefore be understood as a result of raising the surface diffusion. Voids between columns become filled and this causes the sharp decrease of the resistance. Obviously, the more material is melted and the farther the state of the system from equilibrium, the more grain contacts are formed and the deeper the film resistance falls. Different width of the peaks can be connected with wider distribution of nonequilibrium states in the films, prepared at higher deposition rates. The minimum on the resistance curve is observed at temperature 200~K, which is close to $0.3T_m$ ($T_m$(Pb)=601~K). This value agrees with the transition temperature from Zone 1 to Zone 2 (Z2), previously found empirically for many materials~\cite{SmithBook}. Above this temperature the film morphology is determined by surface diffusion and surface free energies of lead and substrate. Here, it is interesting to note that $T$=200~K has been reported as a melting temperature for lead nanoparticles of 3~nm in diameter~\cite{Coombes72}. Obviously, at this temperature the mobility of lead atoms sharply increases. When atoms can freely diffuse on the surface, they form the more energetically favorable structure. Since lead as many other metals doesn't wet insulating substrates, such as glass, the metal ``droplets'' are formed, forming 3D islands. This fact was experimentally observed for lead thin films, deposited in vacuum on glass substrate at room temperature~\cite{Bhaumik}.

AFM investigations (Fig.~\ref{fig2}) showed that the lead films, for which the conductance onset was achieved, had different structure with broader height distribution due to the formation of grain contacts between aggregated nanoparticles. Electrical measurements, as well as previous results, allow to conclude that such a structure is responsible for the observed sensitivity to ammonia and humidity~\cite{Bochenkov02}. The surface of lead particles exposed to the air undergoes oxidation and it becomes covered by oxide layer with the thickness of about 0.5~nm~\cite{Thuermer02}. In this case the conductance of the film is limited by the grain boundaries, that is referred to as ``grain boundary control'' in metal-oxide gas sensor technology~\cite{Yamazoe92}. In our case the particles are not entirely oxide but have a metallic core, covered by oxide shell.

Applying the theory developed for metal-oxide sensors to such systems we suppose that the chemisorbed oxygen atoms on the surface of lead oxide, which covers lead particles, are negatively charged because of trapping electrons from the material. As a result, the surface space charge layer with low concentration of electrons is formed in oxide shell and resistance of the film is high. Adsorption of water molecules can decrease the resistance via blocking the oxygen adsorption sites and due to reactions with adsorbed oxygen, forming the hydroxyl groups. Similarly, the ammonia molecules can compete with oxygen for adsorption sites or react with it. 

\section{Conclusions}
\label{conclusions}
The growth of vacuum deposited lead thin films on cold insulating substrates was studied. Electrical measurements during the deposition showed the dependence of the percolation threshold on the incident energy of adatoms. The minimum on the resistance curves during the annealing of the films were observed at 200~K. AFM study of the annealed samples indicated that two types of nanostructures were formed, depending on the conditions of the deposition. If the conductance onset had not been achieved during the deposition, the resulting film consisted of isolated lead islands. In the opposite case the nanoparticles were interconnected due to aggregation. The growth model based on ballistic aggregation is proposed. 

The sensitivity to humidity was found only for the samples with interconnected nanoparticles. It is reasonable to suppose that the analogous structure is responsible for the chemiresistive properties of lead thin films, grown on poly(\emph{p}-xylylene). 

\ack
The work was financially supported by INTAS grant 2000-00-911 and RFBR grants 02-03-32469 and 03-03-06512. Special thanks to DAAD for the equipment grant.

% \appendix

\bibliographystyle{ELSART-NUM}
\bibliography{GROWTH}

\end{document}